\documentclass[10pt,a4paper]{article} 
\usepackage[margin=1.5cm]{geometry}
\usepackage[numbers]{natbib}
\usepackage{amsmath}
\usepackage{paralist}
\usepackage[hidelinks]{hyperref}
\hypersetup{colorlinks,linkcolor={blue},citecolor={blue},urlcolor={blue}}
\usepackage{graphicx}
\DeclareGraphicsExtensions{.pdf}
\usepackage{listings}
\usepackage{color}
\usepackage{rotating}
\usepackage{mobi}

\newlength{\arrowbump}
\newlength{\arrowlbump}
\newcommand{\PERTURBATIONa}[1]{\overrightharpoondown{#1}}
\newcommand{\PERTURBATIONx}[1]
{\PERTURBATIONa{\rule{0pt}{\arrowbump}\smash[t]{#1}}}
\newcommand{\PERTURBATION}[1]{\PERTURBATIONx{#1}}

\newcommand{\defoccur}[1]{\textsl{#1}}

\title{Efficient Implementation of a Higher-Order Language with Built-In AD%
\thanks{\textbf{Extended abstract presented at the AD 2016 Conference, Sep 2016, Oxford UK.}}}
\author{\href{http://engineering.purdue.edu/~qobi}{Jeffrey Mark
    Siskind}\footnote{Corresponding Author, School of Electrical and Computer
    Engineering, Purdue University,
    \href{mailto:qobi@purdue.edu}{\texttt{qobi@purdue.edu}}}
  \qquad
  \href{http://barak.pearlmutter.net}{Barak A.
    Pearlmutter}\footnote{Dept of Computer Science, National University of
    Ireland Maynooth,
    \href{mailto:barak@pearlmutter.net}{\texttt{barak@pearlmutter.net}}}}
\date{April 2016}

\newcommand{\ie}{\emph{i.e.},}

\newcommand{\etc}{\emph{etc.}}

\newenvironment{lisp}{\begin{tt}\begin{tabular}{l}}{\end{tabular}\end{tt}}

\definecolor{Red}{rgb}{1,0,0}
\definecolor{Green}{rgb}{0,0.69,0}
\definecolor{Blue}{rgb}{0,0,1}
\newcommand{\red}{\color{Red}}
\newcommand{\green}{\color{Green}}
\newcommand{\blue}{\color{Blue}}
\newcommand{\VLAD}{\textsc{vlad}}
\newcommand{\Stalingrad}{\textsc{Stalin$\nabla$}}
\newcommand{\Fortran}{{\mbox{\sc Fortran}}}
\newcommand{\Matlab}{{\mbox{\sc Matlab}}}
\newcommand{\ADIFOR}{{\mbox{\sc adifor}}}
\newcommand{\Tapenade}{{\mbox{\sc Tapenade}}}
\newcommand{\Clang}{{\mbox{\sc c}}}
\newcommand{\ADIC}{{\mbox{\sc adic}}}
\newcommand{\Cplusplus}{{\mbox{\sc c\symbol{43}\symbol{43}}}}
\newcommand{\ADOLC}{{\mbox{\sc adol--c}}}
\newcommand{\CppAD}{\textsc{CppAD}}
\newcommand{\FADBADplusplus}{{\mbox{\sc fadbad\symbol{43}\symbol{43}}}}
\newcommand{\ML}{{\mbox{\sc ml}}}
\newcommand{\MLton}{{\mbox{\sc MLton}}}
\newcommand{\OCaml}{{\mbox{\sc OCaml}}}
\newcommand{\SMLNJ}{{\mbox{\sc sml/nj}}}
\newcommand{\Haskell}{{\mbox{\sc Haskell}}}
\newcommand{\GHC}{{\mbox{\sc ghc}}}
\newcommand{\Scheme}{{\mbox{\sc Scheme}}}
\newcommand{\Bigloo}{{\mbox{\sc Bigloo}}}
\newcommand{\Chicken}{{\mbox{\sc Chicken}}}
\newcommand{\Gambit}{{\mbox{\sc Gambit}}}
\newcommand{\Ikarus}{{\mbox{\sc Ikarus}}}
\newcommand{\Larceny}{{\mbox{\sc Larceny}}}
\newcommand{\MITScheme}{{\mbox{\sc MIT~Scheme}}}
\newcommand{\MzC}{{\mbox{\sc MzC}}}
\newcommand{\MzScheme}{{\mbox{\sc MzScheme}}}
\newcommand{\SchemeToC}{{\mbox{\sc Scheme-$\symbol{62}$C}}}
\newcommand{\SCMUTILS}{\mbox{\textsc{scmutils}}}
\newcommand{\Stalin}{{\mbox{\sc Stalin}}}

\definecolor{darkblue}{rgb}{0,0,0.7}
\definecolor{darkgreen}{rgb}{0,0.5,0}
\definecolor{darkred}{rgb}{0.7,0,0}
\begin{document}
\maketitle
\thispagestyle{empty}

\noindent
We show that AD operators can be provided in a dynamic language without sacrificing numeric performance.
To achieve this, general forward and reverse AD functions are added to a simple high-level dynamic language, and support for them is included in an aggressive optimizing compiler.
Novel technical mechanisms are discussed, which have the ability to migrate the AD transformations from run-time to compile-time.
The resulting system, although only a research prototype, exhibits startlingly good performance.
In fact, despite the potential inefficiencies entailed by support of a functional-programming language and a first-class AD operator, performance is competitive with the fastest available preprocessor-based Fortran AD systems.
On benchmarks involving nested use of the AD operators, it can even dramatically exceed their performance.

\subsection*{The Problem}

Numerical programmers face a tradeoff.
They can use a high-level language, like \Matlab\ or Python, that provides convenient
access to mathematical abstractions like function optimization and
differential equation solvers, or they can use a low-level language, like
\Fortran, to achieve high computational performance.
The convenience of high-level languages results in part from the fact that
they support many forms of run-time dependent computation: storage allocation
and automatic reclamation, data structures whose size is run-time dependent,
pointer indirection, closures, indirect function calls, tags and tag
dispatching, \etc.
This comes at a cost to the numerical programmer: the instruction stream
contains a mix of floating-point instructions and instructions that form the
scaffolding that supports run-time dependent computation.
\Fortran\ code, in contrast, achieves high floating-point performance by avoiding
dilution of the instruction stream with such scaffolding.

This tradeoff is particularly poignant in the domain of automatic
differentiation.
Since the derivative \emph{is} a higher-order function, it is most naturally
incorporated into a language that supports higher-order functions in general.
But on the other hand, efficiency of AD is often critical.

\subsection*{AD Implementation Strategies}

One approach to AD involves a preprocessor performing a
source-to-source transformation.
In its simplest form, this can be viewed as translating
a function:
\begin{flushleft}
\begin{lisp}
double f(double x) \symbol{123}$\ldots$\symbol{125}\\
\end{lisp}
\end{flushleft}
into:
\begin{flushleft}
\begin{lisp}
struct bundle \symbol{123}double primal; double tangent;\symbol{125};\\
struct bundle f\symbol{95}forward(struct bundle x) \symbol{123}$\ldots$\symbol{125}\\
\end{lisp}
\end{flushleft}
that, when passed a bundle of $x$ and $\PERTURBATION{x}$, returns a bundle of
the primal value $f(x)$ and the tangent value $\PERTURBATION{x} f'(x)$.
When implemented properly, repeated application of this transformation can be
used to produce variants of \texttt{f} that compute higher-order derivatives.
Herein lies the inconvenience of this approach.
Different optimizers might use derivatives of different order.
Changing code to use a different optimizer would thus entail changing the build
process to transform the objective function a different number of times.
Moreover, the build process for nested application, such as multilevel
optimization, would be tedious.
One would need to transform the inner objective function, wrap it in a call to
\texttt{optimize}, and then transform this resulting outer function.

\subsection*{A High-Performance Testbed Dynamic Language and Aggressive Compiler}

We present a powerful and expressive formulation of forward and reverse AD based on a novel
set of higher-order primitives, and develop the novel implementation techniques
necessary to support highly efficient implementation of this formulation.
We demonstrate that it is possible to combine the speed of \Fortran\ with the
expressiveness of a higher-level functional-programming language
\emph{augmented} with first-class AD.\@

We exhibit a small but powerful language that provides a
mechanism for defining a \texttt{derivative} operator that offers the
convenience of the first approach with the efficiency of the second approach.
This mechanism is formulated in terms of run-time reflection on the body
of~\texttt{f}, when computing \texttt{(derivative~f)}, to transform it into
something like \texttt{f\symbol{95}forward}.
An optimizing compiler then uses whole-program inter-procedural flow analysis
to eliminate such run-time reflection, as well as all other run-time
scaffolding, yielding numerical code with \Fortran-like (or super-\Fortran) efficiency.

These results are achieved by combining
\begin{inparaenum}[(a)]
\item a novel formulation of forward and reverse AD
in terms of a run-time reflexive mechanism that supports first-class nestable
nonstandard interpretation with
\item the migration of the
nonstandard interpretation to compile-time by whole-program
inter-procedural flow analysis.
\end{inparaenum}

It should be noted that the implementation techniques invented for this purpose are, in principle, compatible with the optimizing compilers for procedural languages like Fortran.  In other words, these techniques could be used to add in-language AD constructs to an aggressive optimizing Fortran or C compiler.  In fact, a proof-of-concept has been exhibited which uses a small subset of these methods to build a Fortran AD pre-preprocessor which accepts a dialect of Fortran with in-language AD block constructs and which allows EXTERNAL FUNCTION arguments, and rewrites/expands the code, generating pure Fortran, and then uses existing tools like Tapenade to perform the required AD \citep{Radul-etal-2012a, Radul-etal-2012b}.

\subsection*{Sketch of Implementation Technology}

We present a novel approach that attains the advantages of both the
overloading and transformation approaches.
We define a novel functional-programming language, \VLAD,\footnote{\VLAD\ is
an acronym for \underline{F}unctional \underline{L}anguage for \underline{AD}
with a voiced~F\@.}
that contains mechanisms for transforming code into new code that computes
derivatives.\footnote{This differs from previous work on forward AD in a
  functional context \citep{Karczmarczuk2001, pearlmutter-siskind-popl-2007,
  SiskindPearlmutter2008a, manzyuk-etal-amazing-2015} which adopts an overloading approach.}
These mechanisms apply to the source code that is, at least conceptually, part
of closures, and such transformation happens, at least conceptually, at run
time.
Such transformation mechanisms replace the preprocessor, support a
callee-derives programming style where the callee invokes the transformation
mechanisms on closures provided by the caller, and allow the control flow of a
program to determine the transformations needed to compute derivatives of the
requisite order.
Polyvariant flow analysis is then used to migrate the requisite
transformations to compile time.\footnote{Existing transformation-based AD
preprocessors, like \ADIFOR\ and \Tapenade, use inter-procedural flow analysis
for different incomparable purposes: not to eliminate run-time
reflection, but to determine which subroutines to transform and which variables
need tangents.}

We present a compiler that generates \Fortran-like target code from a class of
programs written in a higher-order functional-programming language with a
first-class derivative operator.
Our compiler uses whole-program inter-procedural flow analysis to drive a code
generator.
Our approach to flow analysis differs from that typically used when generating
non-\Fortran-like code.
First, it is \defoccur{polyvariant}.
Monovariant flow analyses like 0-CFA \citep{Shivers88} are unable to
specialize higher-order functions.
Polyvariant flow analysis is needed to do so.
The need for polyvariant flow analysis is heightened in the presence of a
higher-order derivative operator, \ie\ one that maps functions to their
derivatives.
Second, it is \defoccur{union free}.
The absence of unions in the abstract interpretation supports generation of
code without tags and tag dispatching.
The further absence of recursion in the abstract interpretation means that all
aggregate data will have fixed size and shape that can be determined by flow
analysis allowing the code generator to use unboxed representations without
indirection in data access or run-time allocation and reclamation.
The polyvariant analysis determines the target of all call sites allowing the
code generator to use direct function calls exclusively.
This, combined with aggressive inlining, results in inlined arithmetic
operations, even when such operations are performed by
(overloaded) function calls.
The polyvariant analysis unrolls finite instances of what is written
as recursive data structures.
This, combined with aggressive unboxing, eliminates essentially all
manipulation of aggregate data, including closures.
Our limitation to union-free analyses and finite unrolling of recursive data
structures is not as severe a limitation as it may seem.
The main limitation relative to \Fortran-like code is that we currently do not
support arrays, though this restriction is easily lifted.
Finally, the polyvariant analysis performs finite instances of reflection,
migrating such reflective access to and creation of code from run time to
compile time.
This last aspect of our flow analysis is novel and crucial.
Our novel AD primitives perform source-to-source transformation \emph{within}
the programming language rather than by a preprocessor, by reflective access to
the code associated with closures and the creation of new closures with
transformed code.
Our flow analysis partially evaluates applications of the AD primitives, thus
migrating such reflective access and transformation of code to compile time.

\centerline{\fbox{
\parbox{0.98\columnwidth}{
\textbf{\quad For those who are not compiler experts}, an intuition for these techniques would be that hunks of code, including both entire procedures and smaller code blocks within procedures, are duplicated to make specialized versions for the different arguments that actually occur.
For example, if an optimization routine \texttt{argmax} is called from two different places in the program using two different objective functions, two versions of \texttt{argmax} would be generated, each specialized to one of the objective functions.
If \texttt{argmax} takes a gradient of the objective function it is passed, each of these two versions of \texttt{argmax} would have a known objective function, which would allow the AD transformation of that function to be migrated to compile time.
Similar machinations allow much of the scaffolding around the numeric computation to be removed.
}}}

The effectiveness of these methods at attaining high floating point performance should be apparent from the benchmarking results shown in Figure~\ref{fig:run-one}.

\begin{sidewaysfigure}
  \begin{center}
    \resizebox{\linewidth}{!}{\begin{tabular}{ll|rrrr|rrrr|rr|rr|rrr}
&&\multicolumn{4}{|c}{}&\multicolumn{4}{|c}{}&\multicolumn{2}{|l}{\texttt{probabilistic-}}&\multicolumn{2}{|l}{\texttt{probabilistic-}}&\multicolumn{3}{|c}{}\\
&&\multicolumn{4}{|c}{\texttt{particle}}&\multicolumn{4}{|c}{\texttt{saddle}}&\multicolumn{2}{|l}{\texttt{lambda-calculus}}&\multicolumn{2}{|l}{\texttt{prolog}}&\multicolumn{3}{|c}{\texttt{backprop}}\\
&&\multicolumn{1}{|c}{\texttt{FF}}&\multicolumn{1}{c}{\texttt{FR}}&\multicolumn{1}{c}{\texttt{RF}}&\multicolumn{1}{c}{\texttt{RR}}&\multicolumn{1}{|c}{\texttt{FF}}&\multicolumn{1}{c}{\texttt{FR}}&\multicolumn{1}{c}{\texttt{RF}}&\multicolumn{1}{c}{\texttt{RR}}&\multicolumn{1}{|c}{\texttt{F}}&\multicolumn{1}{c}{\texttt{R}}&\multicolumn{1}{|c}{\texttt{F}}&\multicolumn{1}{c}{\texttt{R}}&\multicolumn{1}{|c}{\texttt{F}}&\multicolumn{1}{c}{\texttt{Fv}}&\multicolumn{1}{c}{\texttt{R}}\\
\hline
\VLAD
&\Stalingrad
&     1.00
&     1.00
&     1.00
&     1.00
&     1.00
&     1.00
&     1.00
&     1.00
&     1.00
&     1.00
&     1.00
&     1.00
&     1.00
&\multicolumn{1}{c}{{\green\rule{1ex}{1ex}}}
&     1.00
\\
\hline
\Fortran
&{\blue \ADIFOR}
&     2.05
&\multicolumn{1}{c}{{\blue\rule{1ex}{1ex}}}
&\multicolumn{1}{c}{{\blue\rule{1ex}{1ex}}}
&\multicolumn{1}{c|}{{\blue\rule{1ex}{1ex}}}
&     5.44
&\multicolumn{1}{c}{{\blue\rule{1ex}{1ex}}}
&\multicolumn{1}{c}{{\blue\rule{1ex}{1ex}}}
&\multicolumn{1}{c|}{{\blue\rule{1ex}{1ex}}}
&\multicolumn{1}{c}{{\green\rule{1ex}{1ex}}}
&\multicolumn{1}{c|}{{\blue\rule{1ex}{1ex}}}
&\multicolumn{1}{c}{{\green\rule{1ex}{1ex}}}
&\multicolumn{1}{c|}{{\blue\rule{1ex}{1ex}}}
&    15.51
&     3.35
&\multicolumn{1}{c}{{\blue\rule{1ex}{1ex}}}
\\

&{\blue \Tapenade}
&     5.51
&\multicolumn{1}{c}{{\red\rule{1ex}{1ex}}}
&\multicolumn{1}{c}{{\green\rule{1ex}{1ex}}}
&\multicolumn{1}{c|}{{\red\rule{1ex}{1ex}}}
&     8.09
&\multicolumn{1}{c}{{\red\rule{1ex}{1ex}}}
&\multicolumn{1}{c}{{\green\rule{1ex}{1ex}}}
&\multicolumn{1}{c|}{{\red\rule{1ex}{1ex}}}
&\multicolumn{1}{c}{{\green\rule{1ex}{1ex}}}
&\multicolumn{1}{c|}{{\green\rule{1ex}{1ex}}}
&\multicolumn{1}{c}{{\green\rule{1ex}{1ex}}}
&\multicolumn{1}{c|}{{\green\rule{1ex}{1ex}}}
&    14.97
&     5.97
&     6.86
\\
\hline
\Clang
&{\blue \ADIC}
&\multicolumn{1}{c}{{\red\rule{1ex}{1ex}}}
&\multicolumn{1}{c}{{\red\rule{1ex}{1ex}}}
&\multicolumn{1}{c}{{\red\rule{1ex}{1ex}}}
&\multicolumn{1}{c|}{{\red\rule{1ex}{1ex}}}
&\multicolumn{1}{c}{{\red\rule{1ex}{1ex}}}
&\multicolumn{1}{c}{{\red\rule{1ex}{1ex}}}
&\multicolumn{1}{c}{{\red\rule{1ex}{1ex}}}
&\multicolumn{1}{c|}{{\red\rule{1ex}{1ex}}}
&\multicolumn{1}{c}{{\green\rule{1ex}{1ex}}}
&\multicolumn{1}{c|}{{\blue\rule{1ex}{1ex}}}
&\multicolumn{1}{c}{{\green\rule{1ex}{1ex}}}
&\multicolumn{1}{c|}{{\blue\rule{1ex}{1ex}}}
&    22.75
&     5.61
&\multicolumn{1}{c}{{\blue\rule{1ex}{1ex}}}
\\
\hline
\Cplusplus
&{\blue \ADOLC}
&\multicolumn{1}{c}{{\red\rule{1ex}{1ex}}}
&\multicolumn{1}{c}{{\red\rule{1ex}{1ex}}}
&\multicolumn{1}{c}{{\red\rule{1ex}{1ex}}}
&\multicolumn{1}{c|}{{\red\rule{1ex}{1ex}}}
&\multicolumn{1}{c}{{\red\rule{1ex}{1ex}}}
&\multicolumn{1}{c}{{\red\rule{1ex}{1ex}}}
&\multicolumn{1}{c}{{\red\rule{1ex}{1ex}}}
&\multicolumn{1}{c|}{{\red\rule{1ex}{1ex}}}
&\multicolumn{1}{c}{{\green\rule{1ex}{1ex}}}
&\multicolumn{1}{c|}{{\green\rule{1ex}{1ex}}}
&\multicolumn{1}{c}{{\green\rule{1ex}{1ex}}}
&\multicolumn{1}{c|}{{\green\rule{1ex}{1ex}}}
&    12.16
&     5.79
&    32.77
\\

&{\blue \CppAD}
&\multicolumn{1}{c}{{\red\rule{1ex}{1ex}}}
&\multicolumn{1}{c}{{\red\rule{1ex}{1ex}}}
&\multicolumn{1}{c}{{\red\rule{1ex}{1ex}}}
&\multicolumn{1}{c|}{{\red\rule{1ex}{1ex}}}
&\multicolumn{1}{c}{{\red\rule{1ex}{1ex}}}
&\multicolumn{1}{c}{{\red\rule{1ex}{1ex}}}
&\multicolumn{1}{c}{{\red\rule{1ex}{1ex}}}
&\multicolumn{1}{c|}{{\red\rule{1ex}{1ex}}}
&\multicolumn{1}{c}{{\green\rule{1ex}{1ex}}}
&\multicolumn{1}{c|}{{\green\rule{1ex}{1ex}}}
&\multicolumn{1}{c}{{\green\rule{1ex}{1ex}}}
&\multicolumn{1}{c|}{{\green\rule{1ex}{1ex}}}
&    54.74
&\multicolumn{1}{c}{{\blue\rule{1ex}{1ex}}}
&    29.24
\\

&{\blue \FADBADplusplus}
&    93.32
&\multicolumn{1}{c}{{\green\rule{1ex}{1ex}}}
&\multicolumn{1}{c}{{\green\rule{1ex}{1ex}}}
&\multicolumn{1}{c|}{{\green\rule{1ex}{1ex}}}
&    60.67
&\multicolumn{1}{c}{{\green\rule{1ex}{1ex}}}
&\multicolumn{1}{c}{{\green\rule{1ex}{1ex}}}
&\multicolumn{1}{c|}{{\green\rule{1ex}{1ex}}}
&\multicolumn{1}{c}{{\green\rule{1ex}{1ex}}}
&\multicolumn{1}{c|}{{\green\rule{1ex}{1ex}}}
&\multicolumn{1}{c}{{\green\rule{1ex}{1ex}}}
&\multicolumn{1}{c|}{{\green\rule{1ex}{1ex}}}
&   132.31
&    46.01
&    60.71
\\
\hline
\ML
&\MLton
&    78.13
&   111.27
&    45.95
&    32.57
&   114.07
&   146.28
&    12.27
&    10.58
&   129.11
&   114.88
&   848.45
&   507.21
&    95.20
&\multicolumn{1}{c}{{\green\rule{1ex}{1ex}}}
&    39.90
\\

&\OCaml
&   217.03
&   415.64
&   352.06
&   261.38
&   291.26
&   407.67
&    42.39
&    50.21
&   249.40
&   499.43
&  1260.83
&  1542.47
&   202.01
&\multicolumn{1}{c}{{\green\rule{1ex}{1ex}}}
&   156.93
\\

&\SMLNJ
&   153.01
&   226.84
&   270.63
&   192.13
&   271.84
&   299.76
&    25.66
&    23.89
&   234.62
&   258.53
&  2505.59
&  1501.17
&   181.93
&\multicolumn{1}{c}{{\green\rule{1ex}{1ex}}}
&   102.89
\\
\hline
\Haskell
&\GHC
&   209.44
&\multicolumn{1}{c}{{\green\rule{1ex}{1ex}}}
&\multicolumn{1}{c}{{\green\rule{1ex}{1ex}}}
&\multicolumn{1}{c|}{{\green\rule{1ex}{1ex}}}
&   247.57
&\multicolumn{1}{c}{{\green\rule{1ex}{1ex}}}
&\multicolumn{1}{c}{{\green\rule{1ex}{1ex}}}
&\multicolumn{1}{c|}{{\green\rule{1ex}{1ex}}}
&\multicolumn{1}{c}{{\green\rule{1ex}{1ex}}}
&\multicolumn{1}{c|}{{\green\rule{1ex}{1ex}}}
&\multicolumn{1}{c}{{\green\rule{1ex}{1ex}}}
&\multicolumn{1}{c|}{{\green\rule{1ex}{1ex}}}
&\multicolumn{1}{c}{{\green\rule{1ex}{1ex}}}
&\multicolumn{1}{c}{{\green\rule{1ex}{1ex}}}
&\multicolumn{1}{c}{{\green\rule{1ex}{1ex}}}
\\
\hline
\Scheme
&\Bigloo
&   627.78
&   855.70
&   275.63
&   187.39
&  1004.85
&  1076.73
&   105.24
&    89.23
&   983.12
&  1016.50
& 12832.92
&  7918.21
&   743.26
&\multicolumn{1}{c}{{\green\rule{1ex}{1ex}}}
&   360.07
\\

&\Chicken
&  1453.06
&  2501.07
&   821.37
&  1360.00
&  2276.69
&  2964.02
&   225.73
&   252.87
&  2324.54
&  3040.44
& 44891.04
& 24634.44
&  1626.73
&\multicolumn{1}{c}{{\green\rule{1ex}{1ex}}}
&  1125.24
\\

&\Gambit
&   578.94
&   879.39
&   356.47
&   260.98
&   958.73
&  1112.70
&    89.99
&    89.23
&  1033.46
&  1107.26
& 26077.48
& 14262.70
&   671.54
&\multicolumn{1}{c}{{\green\rule{1ex}{1ex}}}
&   379.63
\\

&\Ikarus
&   266.54
&   386.21
&   158.63
&   116.85
&   424.75
&   527.57
&    41.27
&    42.34
&   497.48
&   517.89
&  8474.57
&  4845.10
&   279.59
&\multicolumn{1}{c}{{\green\rule{1ex}{1ex}}}
&   165.16
\\

&\Larceny
&   964.18
&  1308.68
&   360.68
&   272.96
&  1565.53
&  1508.39
&   126.44
&   112.82
&  1658.27
&  1606.44
& 25411.62
& 14386.61
&  1203.34
&\multicolumn{1}{c}{{\green\rule{1ex}{1ex}}}
&   511.54
\\

&\MITScheme
&  2025.23
&  3074.30
&   790.99
&   609.63
&  3501.21
&  3896.88
&   315.17
&   295.67
&  4130.88
&  3817.57
& 87772.39
& 49814.12
&  2446.33
&\multicolumn{1}{c}{{\green\rule{1ex}{1ex}}}
&  1113.09
\\

&\MzC
&  1243.08
&  1944.00
&   740.31
&   557.45
&  2135.92
&  2434.05
&   194.49
&   187.53
&  2294.93
&  2346.13
& 57472.76
& 31784.38
&  1318.60
&\multicolumn{1}{c}{{\green\rule{1ex}{1ex}}}
&   754.47
\\

&\MzScheme
&  1309.82
&  1926.77
&   712.97
&   555.28
&  2371.35
&  2690.64
&   224.61
&   219.29
&  2721.35
&  2625.21
& 60269.37
& 33135.06
&  1364.14
&\multicolumn{1}{c}{{\green\rule{1ex}{1ex}}}
&   772.10
\\

&\SchemeToC
&   582.20
&   743.00
&   270.83
&   208.38
&   910.19
&   913.66
&    82.93
&    69.87
&   811.37
&   803.22
& 10605.32
&  5935.56
&   597.67
&\multicolumn{1}{c}{{\green\rule{1ex}{1ex}}}
&   280.93
\\

&{\blue \SCMUTILS}
&  4462.83
&\multicolumn{1}{c}{{\blue\rule{1ex}{1ex}}}
&\multicolumn{1}{c}{{\blue\rule{1ex}{1ex}}}
&\multicolumn{1}{c|}{{\blue\rule{1ex}{1ex}}}
&  7651.69
&\multicolumn{1}{c}{{\blue\rule{1ex}{1ex}}}
&\multicolumn{1}{c}{{\blue\rule{1ex}{1ex}}}
&\multicolumn{1}{c|}{{\blue\rule{1ex}{1ex}}}
&  7699.14
&\multicolumn{1}{c|}{{\blue\rule{1ex}{1ex}}}
& 83656.17
&\multicolumn{1}{c|}{{\blue\rule{1ex}{1ex}}}
&  5889.26
&\multicolumn{1}{c}{{\blue\rule{1ex}{1ex}}}
&\multicolumn{1}{c}{{\blue\rule{1ex}{1ex}}}
\\

&\Stalin
&   364.08
&   547.73
&   399.39
&   295.00
&   543.68
&   690.64
&    63.96
&    52.93
&   956.47
&  1994.44
& 15048.42
& 16939.28
&   435.82
&\multicolumn{1}{c}{{\green\rule{1ex}{1ex}}}
&   281.27
\\
\end{tabular}
}
  \end{center}
  \caption{Comparative benchmark results for the \texttt{particle} and
    \texttt{saddle} examples \citep{tr-ece-08-01}, the
    \texttt{probabilistic-lambda-calculus} and
    \texttt{probabilistic-prolog} examples \citep{siskind2008} and an
    implementation of backpropagation in neural networks using AD.\@
    Column labels are for AD modes and nesting: F for forward, Fv for
    forward-vector aka stacked tangents, RF for reverse-over-forward,
    etc.
    All run times normalized relative to a unit run time for \Stalingrad\ on
    the corresponding example except that run times for \texttt{backprop-Fv}
    are normalized relative to a unit run time for \Stalingrad\ on
    \texttt{backprop-F}.
    Pre-existing AD tools are named in {\blue blue}, others are custom
    implementations.
    Key:
    {\green\protect\rule{1ex}{1ex}}~not implemented but could implement,
    including \Fortran, \Clang, and \Cplusplus;
    {\blue\protect\rule{1ex}{1ex}}~not implemented in pre-existing AD tool;
    {\red\protect\rule{1ex}{1ex}}~problematic to implement.
    All code available at
    \url{http://www.bcl.hamilton.ie/~qobi/ad2016-benchmarks/}.
  }
  \label{fig:run-one}
\end{sidewaysfigure}

\subsection*{Novelty and Significance}
\label{sec:novelty}

This paper makes two specific novel contributions:
\begin{compactenum}[(1)]
\item A novel set of higher-order functions (\texttt{j*}, \texttt{primal},
\texttt{tangent}, \texttt{bundle}, \texttt{zero}, \texttt{*j}) for performing forward and reverse
AD in a functional language using source-to-source transformation via run-time
reflection.
\item A novel approach for using polyvariant flow analysis to eliminate such
run-time reflection along with all other non-numerical scaffolding.
\end{compactenum}
These are significant because they support AD with a
programming style that is much more expressive and convenient than
that provided by the existing preprocessor-based source-to-source
transformation approach, yet still provides the performance advantages of that
approach.

An alternative perspective would be that the language discussed here is in simple terms an eager lambda calculus augmented with a numeric basis and an AD basis.
This is isomorphic to the intermediate forms used inside high-performance compilers, including aggressive optimizing \Fortran\ compilers.
As such, the techniques we discuss are useful not only for adding AD operators to functional programming languages, but also for adding AD support to compilers for imperative languages.





\section*{Acknowledgments}
This work was supported, in part, by NSF grant 1522954-IIS and by Science
Foundation Ireland grant 09/IN.1/I2637.
Any opinions, findings, and conclusions or recommendations expressed in this
material are those of the authors and do not necessarily reflect the views
of the sponsors.




\bibliographystyle{unsrtnat}
\begin{small}
  \setlength{\bibsep}{0.2ex}
  \bibliography{ad2016b,../sty/QobiTeX}
\end{small}

\end{document}